\documentclass[pra,nobibnotes,aps,showpieces,superscriptaddress,amsfonts,amsmath,float fix,backend=bibtex]{revtex4}
\usepackage[capitalize]{cleveref}
\usepackage{graphicx}
\usepackage{color}
\usepackage{import}
\usepackage{wrapfig}
\usepackage{mathtools}
\usepackage{bm}
\usepackage{amsmath}
\usepackage{amsfonts}
\usepackage{amssymb}

\begin{document}


\title{Modeling the amplitude and energy decay of a weakly damped harmonic oscillator using the energy dissipation rate and a simple trick}

\def\correspondingauthor{\footnote{Corresponding author: klelas@ttf.unizg.hr}}

\author{Karlo Lelas\correspondingauthor{}}
\affiliation{Faculty of Textile Technology, University of Zagreb, Prilaz baruna Filipovića 28a, 10000 Zagreb, Croatia}

\author{Robert Pezer}
\affiliation{Department of Physical Metallurgy, Faculty of Metallurgy, University of Zagreb, Aleja narodnih heroja 3, 44000 Sisak, Croatia}

\date{\today}

\begin{abstract}
We demonstrate how to derive the exponential decrease of amplitude and an excellent approximation of the energy decay of a weakly damped harmonic oscillator without solving the associated equation of motion and without insight into the analytical form of its solution. This is achieved using a basic understanding of the undamped harmonic oscillator and the connection between the damping force's power and the energy dissipation rate. The trick is adding the energy dissipation rates corresponding to two specific pairs of initial conditions with the same energy. In this way, we obtain a first-order differential equation from which we get the time-dependent amplitude and the energies corresponding to each pair of considered initial conditions. Comparing the results of our model to the exact solutions and energies yielded an excellent agreement. The physical concepts and mathematical techniques we employ are well-known to first-year undergraduates.
\end{abstract}

\maketitle

\section{Introduction}

The damped harmonic oscillator, with damping force linear in velocity, is regularly covered in physics textbooks for the first year of undergraduate studies, e.g., see \cite{Cutnell8, Resnick10, Young2020university}. Sometimes, only figures with a graphic representation of solutions of the corresponding equation of motion are given \cite{Cutnell8}, and occasionally analytical expressions of these solutions are given along with such figures \cite{Resnick10, Young2020university}. Still, none of these books show the derivation of these solutions. The reason for this is that the equation of motion for the damped harmonic oscillator is a second-order differential equation, and finding its solution involves mathematical methods that students at that level have not yet learned. Thus, solutions are just given, and usually, only the exponential approximation of the energy decay in the weak damping limit is discussed \cite{Resnick10, Young2020university}. Furthermore, even in physics textbooks that give a more thorough analysis of vibrations and waves at a more advanced undergraduate level, an analytical form of the damped harmonic oscillator solution is given without derivation and without explanation why the amplitude is assumed to decay exponentially with time, e.g., see \cite{Waves}.

On the other hand, in these same textbooks \cite{Cutnell8, Resnick10, Young2020university}, before getting to the part with the damped harmonic oscillator, the simple harmonic oscillator (i.e., undamped harmonic oscillator) is treated in great detail. Students know the solutions of an undamped harmonic oscillator and how to relate the amplitude and phase of the undamped solution to the initial conditions. Another important concept that students learn even before learning about undamped harmonic motion is the concept of instantaneous power and its connection to the rate of change of the system's energy \cite{Resnick10, Young2020university}. 

In this paper, we show how to model the solutions and the energy of a weakly damped harmonic oscillator, using only prior knowledge about the undamped harmonic oscillator and the connection between the power of the damping force and the energy dissipation rate. Of course, we must also have some physical insight into the behavior of the system we are modeling. Therefore, it would be best to start by providing students with figures of the results of carefully conducted experiments on systems described by the damped harmonic oscillator model, such as experiments with mechanical oscillatory systems \cite{Wang, Hinrichse2019, Gonzalez2006} or with an oscillating RLC circuits \cite{Lelas2023}, in a weakly damped regime. Another option we choose in this paper is to show graphically the exact solutions of the equation of motion of a weakly damped harmonic oscillator as a substitute for the experimental results to gain insight into its behavior and the approximations we can use.  

Exponential amplitude decay and purely exponential approximation of the energy decay of weakly damped harmonic oscillator have been derived using the approximation that the amplitude remains constant over time intervals of one period in length and using the energy dissipation rate averaged over these time intervals \cite{Wang, Kontomaris2024}. Our approach allows the amplitude to vary within the period, i.e., there is no need for time averaging, and we obtain the full behaviour of the energy decay.

This paper is organized into six sections. In section \ref{Basic}, we present the theory needed for modeling a weakly damped harmonic oscillator, and we state the problem we are dealing with in this paper. In section \ref{model}, we explain the approximations we use in our model and derive the exponential decrease of the amplitude and the corresponding expressions for the energies for the two pairs of initial conditions we consider. In section \ref{comparison}, we compare the results of our model with exact solutions and energies. In section \ref{simple}, we comment on a simpler version of our model and derivation, which are more suitable for high school students. In section \ref{benefit}, we comment on the benefits of our approach and give suggestions on how to integrate it into teaching. In section \ref{experiment}, we briefly comment on how to experimentally obtain the data for figures needed in our approach.

\section{Basic theory and the statement of the problem}
\label{Basic}

As an example of a damped harmonic oscillator, we consider a block of mass $m$ that oscillates under the influence of the restoring force of an ideal spring of stiffness $k$ and the damping force $F_d(t)=-bv(t)$, where $b$ is the damping constant and $v(t)$ is the velocity of the block. The equation of motion of this system is \cite{Resnick10}
\begin{equation}
ma(t)=-bv(t)-kx(t)\,,
\label{HOeq}
\end{equation}
where $x(t)$ is the block's displacement from the equilibrium position, and $a(t)$ is its acceleration. Since $v(t)=dx(t)/dt$ and $a(t)=d^2x(t)/dt^2$, \eqref{HOeq} is a second-order linear differential equation. The methods for finding its solutions are not given in physics textbooks for the first year of undergraduate studies, e.g., see \cite{Cutnell8, Resnick10, Young2020university}. For any $b\geq0$, the energy (potential plus kinetic) of the block-spring system is given by  
\begin{equation}
E(t)=\frac{kx^2(t)}{2}+\frac{mv^2(t)}{2}\,.
\label{Energy}
\end{equation}
If we put $b=0$ in \eqref{HOeq} we get the equation of the undamped harmonic oscillator \cite{Resnick10}, with general solution $x(t)=A_0\cos(\omega_0t+\phi_0)$, where $\omega_0=\sqrt{k/m}$ is the angular frequency of the undamped system, while $A_0$ and $\phi_0$ are constants (amplitude and phase) that are determined from the initial conditions. The undamped system oscillates with conserved, i.e., constant, energy \cite{Resnick10}. For the damped systems, i.e., for $b>0$, the energy is not conserved due to the power of the damping force $P_d(t)=F_d(t)v(t)$, and the energy dissipation rate is given by \cite{Young2020university} 
\begin{equation}
\frac{dE(t)}{dt}=P_d(t)=-bv^2(t)\,.
\label{Power}
\end{equation}
Equation \eqref{Power} tells us that the energy of the damped system decreases monotonically with time, i.e. $\frac{dE(t)}{dt}\leq0$ $\forall t$, and $\frac{dE(t)}{dt}=0$ holds at the turning points, i.e. at instants when $v(t)=0$. 

\begin{figure}[h!t!]
\begin{center}
\includegraphics[width=0.48\textwidth]{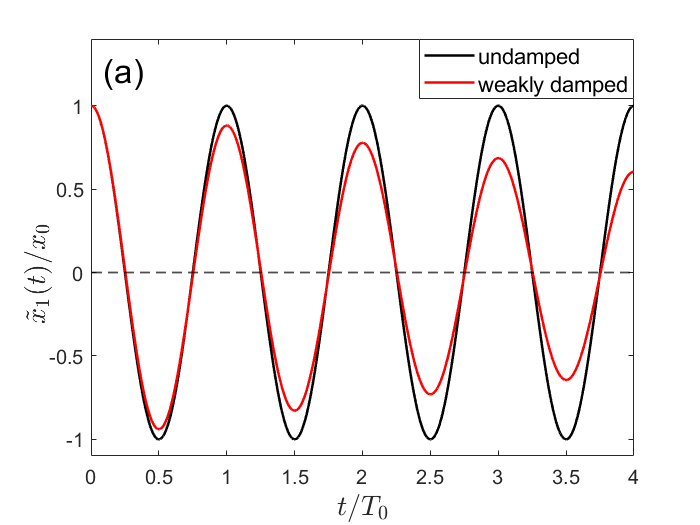}
\includegraphics[width=0.48\textwidth]{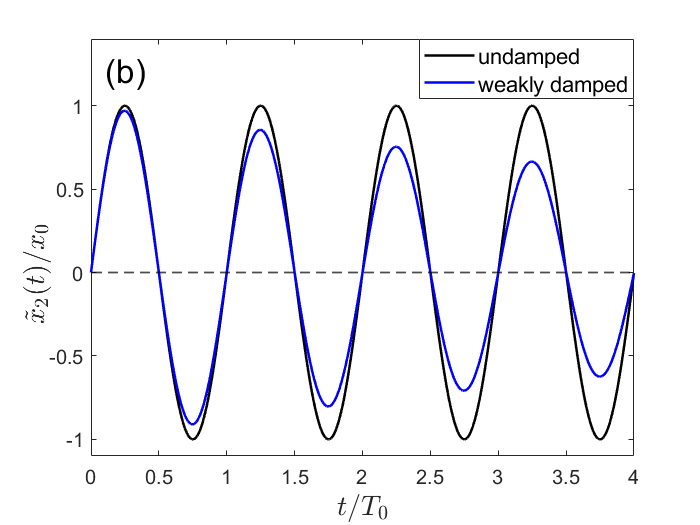}
\end{center}
\caption{(a) The red curve shows the displacement $\tilde{x}_1(t)$ of weakly damped system (with some $m$, $k$ and $b$) for initial conditions $\tilde{x}_1(0)=x_0$ and $\tilde{v}_1(0)=0$. For comparison, the black curve shows the undamped solution with the same initial conditions, i.e., $x(t)=x_0\cos(\omega_0t)$. (b) The blue curve shows the displacement $\tilde{x}_2(t)$ of the same weakly damped system for initial conditions $\tilde{x}_2(0)=0$ and $\tilde{v}_2(0)=v_0=\omega_0x_0$, while the black curve shows the undamped solution with these initial conditions, i.e., $x(t)=x_0\sin(\omega_0t)$, for comparison. In both figures, the time is in $T_0=2\pi/\omega_0$ units, and the displacement is in $x_0$ units.} 
\label{x1x2}
\end{figure}

Let's imagine that the instructor gives his students Fig.\,\ref{x1x2} with graphic representations of the exact solutions of equation \eqref{HOeq} for a weakly damped system and an undamped system, for two pairs of initial conditions, which have the same initial energy $E_0$. One pair of initial conditions has a purely potential initial energy, while the other pair of initial conditions has a purely kinetic initial energy. The analytical expressions of the damped solutions used in Fig.\,\ref{x1x2} are given in section \ref{comparison}. Now comes the statement of the problem. The instructor says to the students: "Find the expression for the energy decay of weakly damped system, using the insights provided by Fig.\,\ref{x1x2} and the energy dissipation rate \eqref{Power}. Furthermore, find what condition $b$ must satisfy, compared to $m$ and $k$, for the system to be weakly damped."

\section{Modeling of the amplitude decrease and the energy decay of weakly damped harmonic oscillator}
\label{model}

In Fig.\,\ref{x1x2}(a) and (b), we see that the frequency and phase of $\tilde{x}_1(t)$ and $\tilde{x}_2(t)$ remain unchanged compared to the undamped solutions (at least as far as we can see from Fig.\,\ref{x1x2}(a) and (b)), while the amplitudes of $\tilde{x}_1(t)$ and $\tilde{x}_2(t)$ slowly decrease with time in the same fashion. It is reasonable to describe the slow decay of the amplitudes of $\tilde{x}_1(t)$ and $\tilde{x}_2(t)$ with the same function, even though we can notice that the first few extremes of $\tilde{x}_2(t)$ are slightly smaller in magnitude than the extremes of $\tilde{x}_1(t)$ (for later times, the difference in the magnitude of extremes is hardly visible), since this can be explained by the fact that $\tilde{x}_2(t)$ reaches each extreme $T_0/4$ time later than $\tilde{x}_1(t)$, so $\tilde{x}_2(t)$ is exposed to damping for more extended time up to the moment of reaching the extremes compared to $\tilde{x}_1(t)$. These insights suggest that the weakly damped displacements shown in Fig.\,\ref{x1x2}(a) and (b) can be modeled with
\begin{equation}
x_1(t)=f(t)x_0\cos(\omega_0 t)\, 
\label{x1}
\end{equation}
and 
\begin{equation}
x_2(t)=f(t)x_0\sin(\omega_0 t)\,
\label{x2}
\end{equation}
where function $f(t)$ describes the amplitude decrease in time. From \eqref{x1} and \eqref{x2} we calculate the corresponding velocities
\begin{equation}
v_1(t)=\frac{dx_1(t)}{dt}=-f(t)\omega_0x_0\sin(\omega_0 t)+\dot{f}(t)x_0\cos(\omega_0 t) 
\label{v1}
\end{equation}
and 
\begin{equation}
v_2(t)=\frac{dx_2(t)}{dt}=f(t)\omega_0x_0\cos(\omega_0 t)+\dot{f}(t)x_0\sin(\omega_0 t)\,,
\label{v2}
\end{equation}
where we used notation $\dot{f}(t)=df(t)/dt$. Since $\tilde{x}_1(0)=x_0$ and $\tilde{v}_2(0)=\omega_0x_0$ we can easily see that $f(0)=1$ must hold in order that our modeled solutions $x_1(t)$ and $x_2(t)$ satisfy these initial conditions. Furthermore, our modeled solution $x_1(t)$ has a initial velocity, i.e. $v_1(0)=\dot{f}(0)x_0$, while $\tilde{x}_1(t)$ has zero initial velocity. We deal with this issue in the following two paragraphs.  

It is clear from Fig.\,\ref{x1x2}(a) and (b) that the amplitudes of $\tilde{x}_1(t)$ and $\tilde{x}_2(t)$ change very slowly over time in comparison to the oscillating parts. Therefore, the rate of change of function $f(t)$ over time has to be much slower than the rate of change of $\cos(\omega_0t)$ and $\sin(\omega_0t)$, which is of the order $\omega_0$. We can conclude that $|\dot{f}(t)|\ll\omega_0$ holds for weak damping for all time, i.e., for $t\geq0$ (we have put absolute value because we can expect the negative derivative of $f(t)$ since the amplitude decreases with time). 

Regarding the discrepancy of our modeled solution $x_1(t)$, i.e., \eqref{x1}, and $\tilde{x}_1(t)$ in initial velocity, we can easily see that adding a phase $\phi$ to cosine in $x_1(t)$ could resolve this issue since it would lead to the initial velocity of the form
\begin{equation}
v_1(0)=-f(0)\omega_0x_0\sin(\phi)+\dot{f}(0)x_0\cos(\phi) 
\label{v1in}
\end{equation}
and, by setting $v_1(0)=0$, one easily gets that $\phi=\arctan(\dot{f}(0)/\omega_0))$ resolves this issue, but, due to the condition $|\dot{f}(0)|\ll\omega_0$, we can see that $|\phi|\ll1$ holds. This leads us to the conclusion that $\tilde{x}_1(t)$ has a phase, but which is so tiny that it is unobservable in Fig.\,\ref{x1x2}(a). Thus, adding this phase into our model would be redundant, i.e., \eqref{x1} is a valid model of $\tilde{x}_1(t)$.      

By using our modeled displacements and velocities in \eqref{Energy}, we get the corresponding energies     
\begin{equation}
E_1(t)=E_0\left[f^2(t)-f(t)\left(\frac{\dot{f}(t)}{\omega_0}\right)\sin(2\omega_0t)+\left(\frac{\dot{f}(t)}{\omega_0}\right)^2\cos^2(\omega_0t)\right] 
\label{E1}
\end{equation}
and 
\begin{equation}
E_2(t)=E_0\left[f^2(t)+f(t)\left(\frac{\dot{f}(t)}{\omega_0}\right)\sin(2\omega_0t)+\left(\frac{\dot{f}(t)}{\omega_0}\right)^2\sin^2(\omega_0t)\right]\,, 
\label{E2}
\end{equation}
where $E_0=m\omega_0^2x_0^2/2$ is the initial energy, and we used well know trigonometric identities $\sin^2(\omega_0t)+\cos^2(\omega_0t)=1$ and $\sin(2\omega_0t)=2\sin(\omega_0t)\cos(\omega_0t)$ in deriving \eqref{E1} and \eqref{E2}. We can see that our modeled solution $x_2(t)$ has the right initial energy, i.e. $E_2(0)=E_0$, while $x_1(t)$ has initial energy $E_1(0)=E_0\left(1+(\dot{f}(0)/\omega_0)^2\right)$. Since $|\dot{f}(t)|\ll\omega_0$ holds for weak damping, we can safely neglect the terms of the order $(\dot{f}(t)/\omega_0)^2$ in \eqref{E1} and \eqref{E2}. Thus, to the first order in $(\dot{f}(t)/\omega_0)$, our modeled expressions for the energies are 
\begin{equation}
E_1(t)=E_0\left[f^2(t)-f(t)\left(\frac{\dot{f}(t)}{\omega_0}\right)\sin(2\omega_0t)\right] 
\label{E11}
\end{equation}
and 
\begin{equation}
E_2(t)=E_0\left[f^2(t)+f(t)\left(\frac{\dot{f}(t)}{\omega_0}\right)\sin(2\omega_0t)\right]\,. 
\label{E22}
\end{equation}
Both \eqref{E11} and \eqref{E22} have the correct initial energy. Now we focus on determining the $f(t)$.

If one would put, e.g., the energy \eqref{E11} into the equation \eqref{Power}, i.e., $dE_1(t)/dt=-bv_1(t)^2$, it would lead to a very complicated differential equation for $f(t)$, but instead of that we can use the following trick. We add the dissipation rates corresponding to two pairs of initial conditions, i.e.  
\begin{equation}
\frac{d}{dt}\left(E_1(t)+E_2(t)\right)=-b\left(v_1^2(t)+v_2^2(t)\right)\,. 
\label{trick}
\end{equation}
Since $E_1(t)+E_2(t)=2E_0f^2(t)$, and $v_1^2(t)+v_2^2(t)=\omega_0^2x_0^2f^2(t)$ (in adding the squared velocities we again neglected the terms of the order $(\dot{f}(t)/\omega_0)^2$) we get the differential equation 
\begin{equation}
\frac{d}{dt}f^2(t)=-\frac{b}{m}f^2(t)\,. 
\label{trick1}
\end{equation}
Using substitutions $F(t)=f^2(t)$ and $\alpha=-b/m$, \eqref{trick1} becomes $dF(t)/dt=\alpha F(t)$, and first year undergraduates know that the function $F(t)=e^{\alpha t}$ satisfies this differential equation \cite{Young2020university}.
Thus, we obtain 
\begin{equation}
f(t)=\sqrt{F(t)}=e^{-\frac{b}{2m}t}
\label{amplitude}
\end{equation}
as the slowly decaying amplitude of our modeled solutions \eqref{x1} and \eqref{x2}. Using \eqref{amplitude}, and its time derivative, in \eqref{E11} and \eqref{E22} we get
\begin{equation}
E_1(t)=E_0e^{-\frac{b}{m}t}\left[1+\frac{b}{2m\omega_0}\sin(2\omega_0t)\right] 
\label{E1f}
\end{equation}
and 
\begin{equation}
E_2(t)=E_0e^{-\frac{b}{m}t}\left[1-\frac{b}{2m\omega_0}\sin(2\omega_0t)\right]\, 
\label{E2f}
\end{equation}
as final expressions for the energies in our model.

We can now easily find what condition $b$ must satisfy, compared to $m$ and $k$, for the system to be weakly damped. As we stated before, for weak damping condition $|\dot{f}(t)|\ll\omega_0$ holds. This condition holds for all $t$, thus, if we consider $t=0$, for which $\dot{f}(0)=-b/(2m)$, we get 
\begin{equation}
\frac{b}{2m}\ll\omega_0
\label{uvjet}
\end{equation}
as the condition of weak damping. Since $\omega_0=\sqrt{k/m}$, \eqref{uvjet} can be rewritten as $b\ll2\sqrt{km}$, which is a common form of the weak damping condition in physics textbooks \cite{Young2020university}.  

We note here that our trick \eqref{trick} of adding the dissipation rates works only for damping force linear in velocity, i.e. for $F_d(t)=-bv(t)$. For other types of damping, such as damping due to sliding friction with $F_d\propto-v(t)/|v(t)|$ \cite{Zonetti}, or damping due to air resistance with $F_d\propto-v(t)^3/|v(t)|$ \cite{Mungan}, our trick does not lead to expressions that can be simplified using trigonometric identities. For all three mentioned types of damping, approximate solutions, in the limit of weak damping, can be obtained by averaging the energy dissipation rate over a period $T_0$ \cite{Wang}, but this method gives only exponential approximation of energy decay in case $F_d(t)=-bv(t)$ and is mathematically more demanding than our method.    

\section{Comparison of our model with exact results}
\label{comparison}

\begin{figure}[h!t!]
\begin{center}
\includegraphics[width=0.48\textwidth]{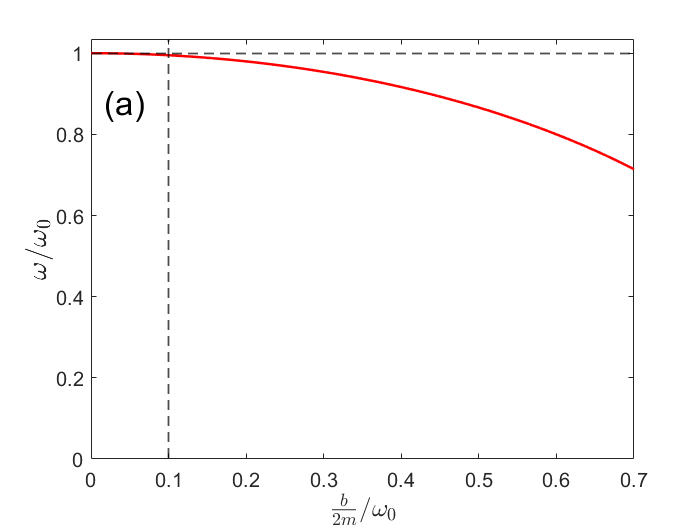}
\includegraphics[width=0.48\textwidth]{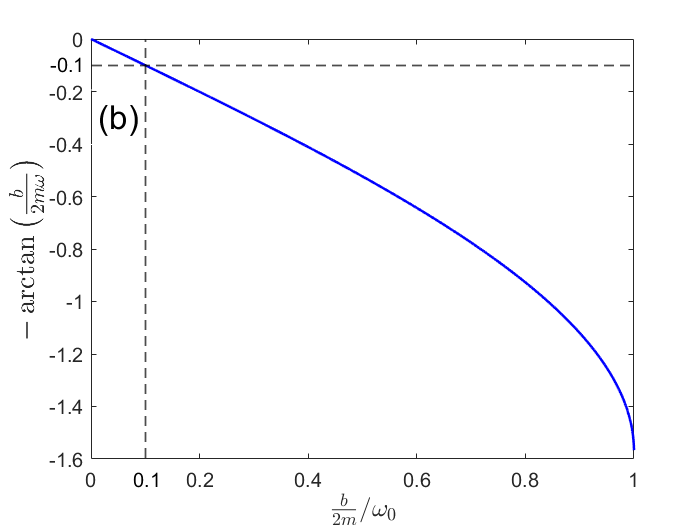}
\end{center}
\caption{(a) Damped angular frequency $\omega$ as a function of ratio $\frac{b}{2m}$. (b) Initial phase of solution \eqref{x1eg} as a function of ratio $\frac{b}{2m}$.} 
\label{Limits}
\end{figure}

The exact solution of differential equation \eqref{HOeq}, for general initial conditions, is given, e.g., in \cite{Lelas2023}. For a pair of initial conditions, which we considered in Fig.\,\ref{x1x2}(a) and (b), the exact solutions are
\begin{equation}
\tilde{x}_1(t)=\frac{\omega_0x_0}{\omega}e^{-\frac{b}{2m}t}\cos\left(\omega t-\arctan\left(\frac{b}{2m\omega}\right)\right)\, 
\label{x1eg}
\end{equation}
and 
\begin{equation}
\tilde{x}_2(t)=\frac{\omega_0x_0}{\omega}e^{-\frac{b}{2m}t}\sin\left(\omega t\right)\, 
\label{x2eg}
\end{equation}
where $\omega=\sqrt{\omega_0^2-\left(\frac{b}{2m}\right)^2}$ is the damped angular frequency. In Fig.\,\ref{x1x2}(a) and (b) we took $\frac{b}{2m}=0.02\omega_0$, which is in good agreement with weak damping condition \eqref{uvjet}. In this case, \eqref{x1eg} and \eqref{x2eg} become
\begin{equation}
\tilde{x}_1(t)=1.0002\,e^{-0.02\omega_0t}x_0\cos\left(0.9998\,\omega_0 t-0.0020\right)\, 
\label{x1eg1}
\end{equation}
and 
\begin{equation}
\tilde{x}_1(t)=1.0002\,e^{-0.02\omega_0t}x_0\sin\left(0.9998\,\omega_0 t\right)\,, 
\label{x2eg2}
\end{equation}
where we rounded numbers to four decimals, thus, even without showing a graphical comparison, it is clear that our modeled solution \eqref{x1} and \eqref{x2}, with $f(t)=e^{-0.02\omega_0t}$, approximate excellently the exact solutions \eqref{x1eg1} and \eqref{x2eg2}. Of course, the agreement between our modeled and exact solutions will be good only if condition \eqref{uvjet} holds because only then $\omega=\omega_0\sqrt{1-\left(\frac{b}{2m\omega_0}\right)^2}\approx\omega_0$ is valid and the initial phase in \eqref{x1eg} is negligible. To quantify the validity of our model, but also to gain insight into its limitations, in Fig.\,\ref{Limits}(a), we show $\omega$ as a function of ratio $\frac{b}{2m}$ and we can see that it starts to significantly deviate from $\omega_0$ at about $\frac{b}{2m}\approx0.1\omega_0$.  
In Fig.\,\ref{Limits}(b), we show the initial phase of the exact solution \eqref{x1eg} as a function of $\frac{b}{2m}$ and we can see that it increases in magnitude more rapidly, with the increase of $\frac{b}{2m}$, than $\omega$ decreases. For $\frac{b}{2m}=0.1\omega_0$, we get $\omega=0.995\omega_0$ and $\arctan\left(\frac{b}{2m\omega}\right)=0.1$, i.e. the first maximum of cosine in exact solution \eqref{x1eg} is shifted forward in time by $\Delta t=0.1/\omega=0.016T_0$, in comparison to the first maximum of cosine in our modeled solution \eqref{x1} at $t=0$. We take these differences in angular frequency and initial phase for $\frac{b}{2m}=0.1\omega_0$ as the limits of acceptable deviation of the modeled solutions from the exact solutions. In that sense, we can safely say that our model describes well the exact solutions for $0<\frac{b}{2m}\lesssim 0.1\omega_0$. 

\begin{figure}[h!t!]
\begin{center}
\includegraphics[width=0.48\textwidth]{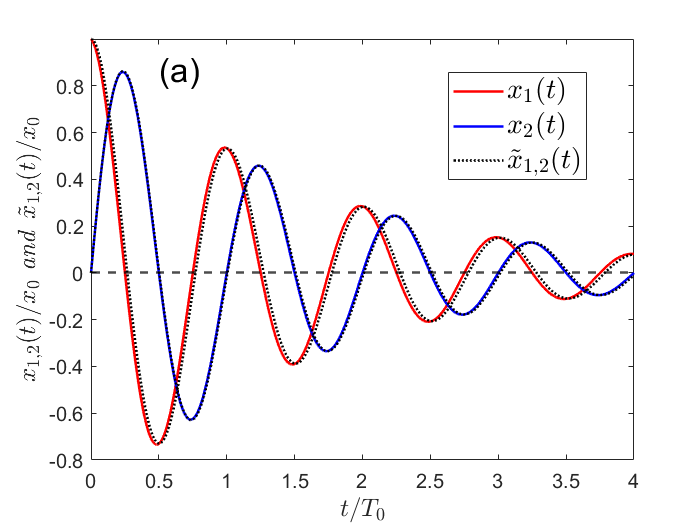}
\includegraphics[width=0.48\textwidth]{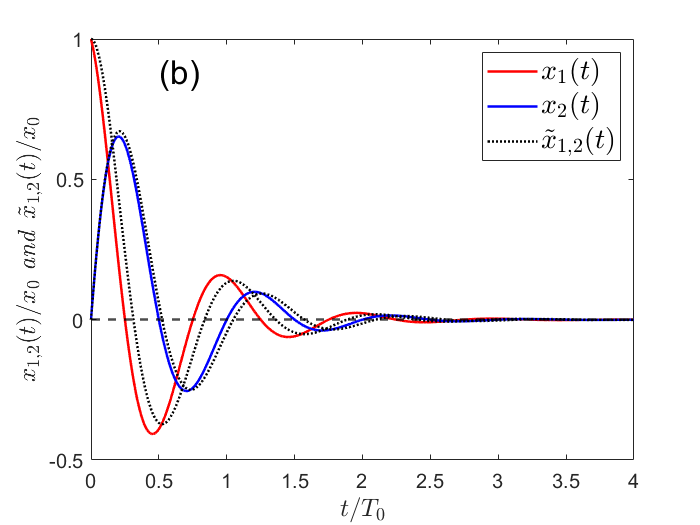}
\end{center}
\caption{Modeled solutions $x_1(t)$ (red curve) and $x_2(t)$ (blue curve) given by \eqref{x1} and \eqref{x2}, for (a) $\frac{b}{2m}=0.1\omega_0$ and (b) $\frac{b}{2m}=0.3\omega_0$. The corresponding exact solutions $\tilde{x}_1(t)$ and $\tilde{x}_2(t)$ given by \eqref{x1eg} and \eqref{x2eg} are shown as dotted curves.} 
\label{xusporedba}
\end{figure}

To back up our choice visually, in Fig.\,\ref{xusporedba}(a) and (b) we show the modeled solutions \eqref{x1} and \eqref{x2} for $\frac{b}{2m}=\lbrace0.1\omega_0,0.3\omega_0\rbrace$, and the corresponding exact solutions \eqref{x1eg} and \eqref{x2eg}. We see only a slight mismatch between the modeled and the exact solutions for $\frac{b}{2m}=0.1\omega_0$, while, as expected, the discrepancy is more pronounced for $\frac{b}{2m}=0.3\omega_0$. Of course, due to the differences in angular frequency, the difference in phase of cosine functions in modeled solutions and exact solutions increases linearly with time, i.e. $\Delta\phi\sim(\omega_0-\omega)t$, but the oscillations, for $0<\frac{b}{2m}\lesssim 0.1\omega_0$, become negligible much sooner than this phase difference significantly increases, since the amplitude decays exponentially with time.     

\begin{figure}[h!t!]
\begin{center}
\includegraphics[width=0.48\textwidth]{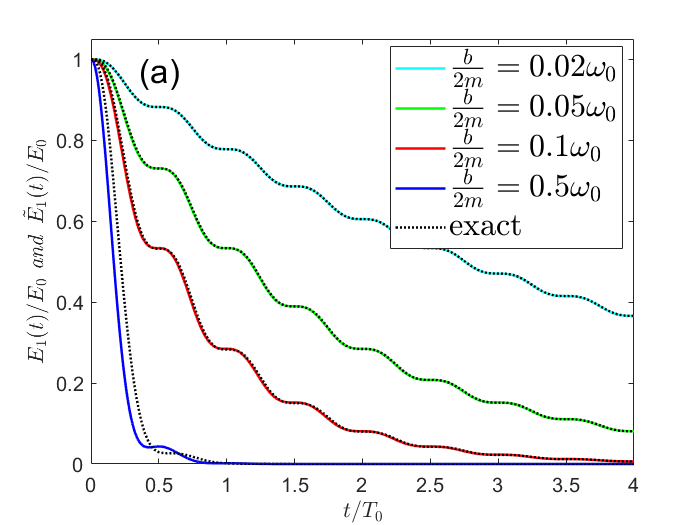}
\includegraphics[width=0.48\textwidth]{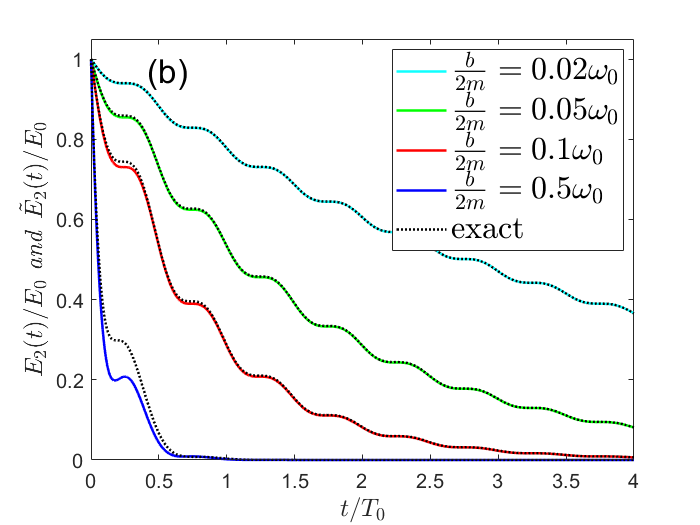}
\end{center}
\caption{(a) Modeled energy $E_1(t)$, i.e. \eqref{E1f}. (b) Modeled energy $E_2(t)$, i.e. \eqref{E2f}. Both figures show the behavior for $\frac{b}{2m}=\lbrace0.02\omega_0,0.05\omega_0,0.1\omega_0,0.5\omega_0\rbrace$. The corresponding exact energies $\tilde{E}_1(t)$ and $\tilde{E}_2(t)$, given by \eqref{Eeg}, are shown as dotted curves on both figures.} 
\label{Eusporedba}
\end{figure}

The exact expression for the energy of the damped harmonic oscillator, for general initial conditions, is given, e.g., in \cite{Lelas2024}. For a pair of initial conditions, which we considered in Fig.\,\ref{x1x2}(a) and (b), the exact energies are
\begin{equation}
\tilde{E}_{1,2}(t)=E_0e^{-\frac{b}{m}t}\left[1\pm\frac{b}{2m\omega}\sin(2\omega t)+2\left(\frac{b}{2m\omega}\right)^2\sin^2(\omega t)\right]\,, 
\label{Eeg}
\end{equation}
where the upper sign in $\pm$ corresponds to $\tilde{E}_1(t)$, and lower sign to $\tilde{E}_2(t)$. Again, it is easy to see that the exact energies reduce to our modeled energies \eqref{E1f} and \eqref{E2f} in case approximations $\omega\approx\omega_0$ and $\left(\frac{b}{2m\omega_0}\right)^2\approx0$ are valid. In Fig.\,\ref{Eusporedba} we show the modeled energies and the exact energies for $\frac{b}{2m}=\lbrace0.02\omega_0,0.05\omega_0,0.1\omega_0, 0.5\omega_0\rbrace$. We can see excellent agreement between our modeled energies and the exact energies for $\frac{b}{2m}=\lbrace0.02\omega_0,0.05\omega_0,0.1\omega_0\rbrace$, with small quantitative differences visible in Fig.\,\ref{Eusporedba}(b) for case $\frac{b}{2m}=0.1\omega_0$, but even in that case the modeled energy seems to show qualitatively good behavior, i.e., it decreases monotonically with plateaus in vicinity of the turning points, as it should according to energy dissipation rate \eqref{Power}. In both Fig.\,\ref{Eusporedba}(a) and (b), we can see that the modeled energies for $\frac{b}{2m}=0.5\omega_0$ show quantitatively and qualitatively wrong behavior, namely local maxima are visible in vicinity of the first turning point in both cases.

One can ask the question, do local maxima in modeled energies \eqref{E1f} and \eqref{E2f} appear for ratios $\frac{b}{2m\omega_0}$ greater than some specific value or are present for all values of that ratio? The first derivatives of modeled energies \eqref{E1f} and \eqref{E2f} are
\begin{equation}
\frac{dE_{1,2}(t)}{dt}=-\left(\frac{b}{m}\right)E_0e^{-\frac{b}{m}t}\left[1\pm\sqrt{1+\left(\frac{b}{2m\omega_0}\right)^2}\sin\left(2\omega_0t-\arctan\left(\frac{2m\omega_0}{b}\right)\right)\right]\,.
\label{E12dissipation}
\end{equation}
In deriving \eqref{E12dissipation} we used known trigonometric identity $a\sin(2\omega_0t)+b\cos(2\omega_0t)=\sqrt{a^2+b^2}\sin(2\omega_0t+\arctan(b/a))$. Since the amplitude of the sine function in \eqref{E12dissipation} is greater than one for any $\frac{b}{2m}>0$, we can easily understand that \eqref{E12dissipation} can be positive near the turning points of modeled solutions \eqref{x1} and \eqref{x2}, i.e. modeled energies \eqref{E1f} and \eqref{E2f} have local maxima in the vicinity of the turning points, but, if the condition \eqref{uvjet} holds, approximation $\left(\frac{b}{2m\omega_0}\right)^2\approx 0$ is valid, and in that case these local maxima are unobservable.      

\begin{figure}[h!t!]
\begin{center}
\includegraphics[width=0.48\textwidth]{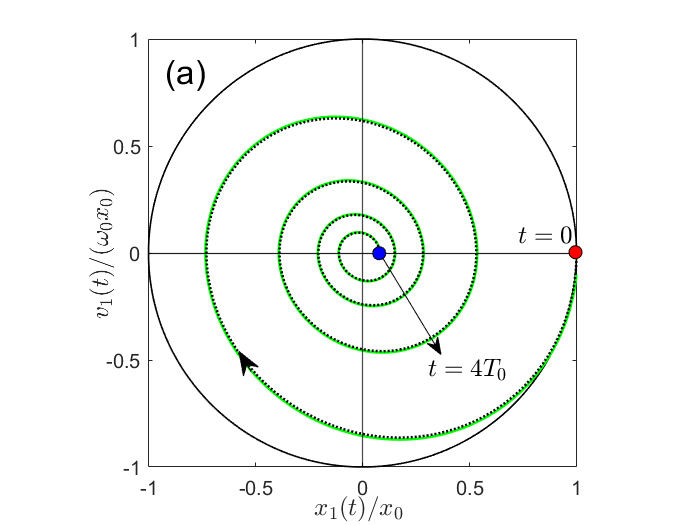}
\includegraphics[width=0.48\textwidth]{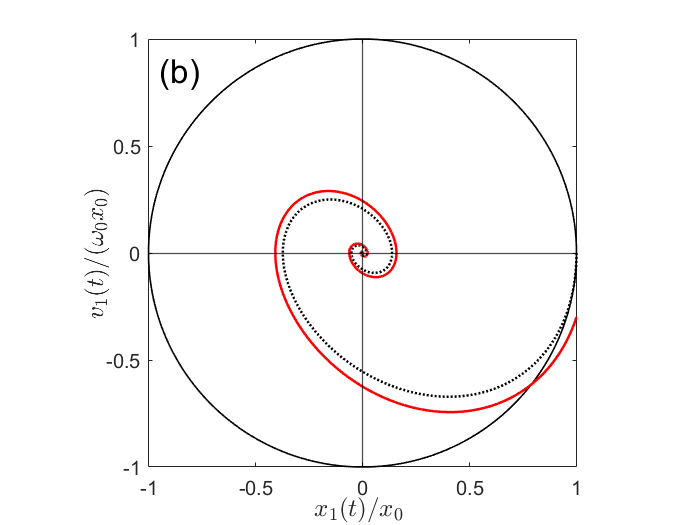}
\end{center}
\caption{Parametric plot of modeled velocity $v_1(t)$, i.e. \eqref{v1}, vs. modeled displacement $x_1(t)$, i.e. \eqref{x1}, with time $t\in[0,4T_0]$ as parameter, shown as a green curve for $\frac{b}{2m}=0.1\omega_0$ in (a), and as a red curve for $\frac{b}{2m}=0.3\omega_0$ in (b). In both figures, the black circles are the corresponding parametric plots of the undamped case, and the black dotted curves are the parametric plots of exact velocities vs. exact displacements for the same initial conditions and the same damping. See text for details.} 
\label{phase1}
\end{figure}
\begin{figure}[h!t!]
\begin{center}
\includegraphics[width=0.48\textwidth]{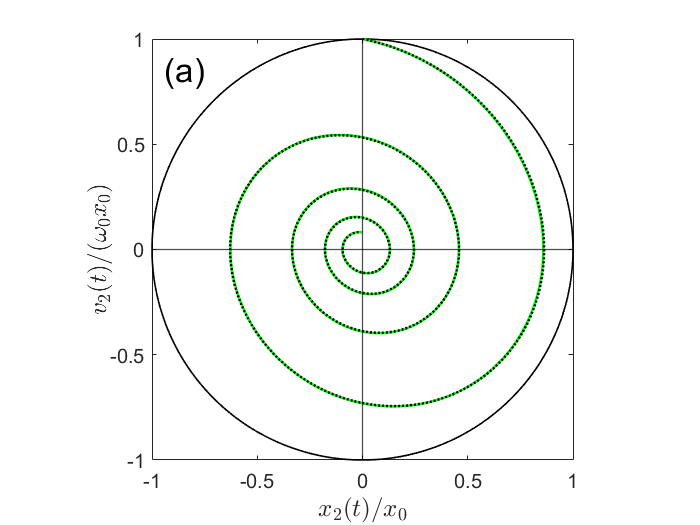}
\includegraphics[width=0.48\textwidth]{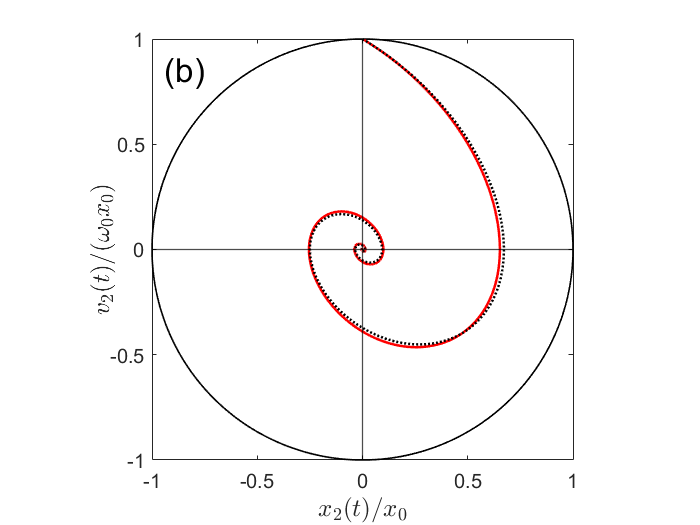}
\end{center}
\caption{Parametric plot of modeled velocity $v_2(t)$, i.e. \eqref{v2}, vs. modeled displacement $x_2(t)$, i.e. \eqref{x2}, with time $t\in[0,4T_0]$ as parameter, shown as a green curve for $\frac{b}{2m}=0.1\omega_0$ in (a), and as a red curve for $\frac{b}{2m}=0.3\omega_0$ in (b). In both figures, the black circles are the corresponding parametric plots of the undamped case, and the black dotted curves are the parametric plots of exact velocities vs. exact displacements for the same initial conditions and the same damping. See text for details.} 
\label{phase2}
\end{figure}

Yet another interesting visualization of the relationship between our modeled solutions and the exact solutions is provided by considering velocity vs. displacement parametric plots (with time as a parameter), which are known as phase space diagrams and are regularly used in classical mechanics \cite{ClassicalM}. In Fig.\,\ref{phase1} we show a parametric plot of modeled velocity $v_1(t)$ plotted as a function of modeled displacement $x_1(t)$ for $\frac{b}{2m}=0.1\omega_0$ (green curve in Fig.\,\ref{phase1}(a)) and for $\frac{b}{2m}=0.3\omega_0$ (red curve in Fig.\,\ref{phase1}(b)). As indicated in Fig.\,\ref{phase1}(a), time flows clockwise, and we consider $0\leq t\leq 4T_0$. In Fig.\,\ref{phase2} we show modeled velocity $v_2(t)$ plotted as a function of modeled displacement $x_2(t)$, for $\frac{b}{2m}=0.1\omega_0$ (green curve in Fig.\,\ref{phase2}(a)) and for $\frac{b}{2m}=0.3\omega_0$ (red curve in Fig.\,\ref{phase2}(b)). On both Fig.\,\ref{phase1} and Fig.\,\ref{phase2}, all dotted black curves show the corresponding exact velocities and exact displacements, i.e. velocities $\tilde{v}_{1,2}(t)=d\tilde{x}_{1,2}(t)/dt$ and displacements $\tilde{x}_{1,2}(t)$. In contrast, black circles show the corresponding parametric plot of the undamped harmonic oscillator. 
Due to the chosen units, the square of the radius of the black circle is equal in magnitude to the energy of the undamped system in $E_0$ units, i.e. equal to one, on all figures. The square of the distance from the origin $(0,0)$ to points $(\tilde{x}_{1,2}(t),\tilde{v}_{1,2}(t))$ on black dotted curves is equal in magnitude to $\tilde{E}_{1,2}(t)$ in $E_0$ units. The square of the distance from the origin $(0,0)$ to points $(x_{1,2}(t),v_{1,2}(t))$ on green and red curves is not equal in magnitude to our modeled energies \eqref{E1f} and \eqref{E2f} in $E_0$ units, but is equal to \eqref{E1} and \eqref{E2} which contain a term of order $\left(\frac{b}{2m\omega_0}\right)^2$. Thus, the square of the distance from the origin to points on the red and green curves is an excellent approximation of our modeled energies \eqref{E1f} and \eqref{E2f} if condition \eqref{uvjet} holds. With all that in mind, Fig.\,\ref{phase1} and Fig.\,\ref{phase2} give us a simultaneous visual comparison of the modeled displacements, velocities and energies with the exact displacements, velocities and energies.  In Fig.\,\ref{phase1}, as noted earlier, we see that our modeled solution $x_1(t)$ does not agree with the exact solution in the initial velocity, i.e. the modeled solution $x_1(t)$ has initial velocity $v_1(0)=-\frac{bx_0}{2m}$, which is clearly visible for $\frac{b}{2m}=0.3\omega_0$. In Fig.\,\ref{phase1} and Fig.\,\ref{phase2}, another interesting feature of the damped harmonic oscillator can be easily seen: the velocity does not have maxima in magnitude when the system passes through the equilibrium position, as it has in the undamped case. This fact has already been pointed out, e.g., in \cite{Hinrichse2019, Lelas2023}.   

\section{A simpler approach}
\label{simple}

Exponential amplitude decay and purely exponential approximation of the energy decay of weakly damped harmonic oscillator have recently been derived, in a manner suitable for high school students, using the energy dissipation rate averaged over one period of the corresponding undamped system, i.e. $T_0$, with the approximation that the amplitude remains constant over time intervals $\Delta t\sim T_0$ \cite{Kontomaris2024}. In our model, this approximation corresponds to neglecting terms with $\dot{f}(t)$ in the velocities \eqref{v1} and \eqref{v2}, while the displacements \eqref{x1} and \eqref{x2} remain the same. In this case, it is easy to show that the modeled energies for both pairs of initial conditions are the same, i.e. 
\begin{equation}
E_1(t)=E_2(t)=\frac{m\omega_0^2x_0^2}{2}f^2(t)\,, 
\label{Esimple}
\end{equation}
and the corresponding dissipation rates \eqref{Power} are 
\begin{equation}
\frac{dE_1(t)}{dt}=-bf^2(t)\omega_0^2x_0^2\sin^2(\omega_0t)\, 
\label{E1simple}
\end{equation}
and
\begin{equation}
\frac{dE_2(t)}{dt}=-bf^2(t)\omega_0^2x_0^2\cos^2(\omega_0t)\,. 
\label{E2simple}
\end{equation}
Now, it is even easier to see that the sum of \eqref{E1simple} and \eqref{E2simple} leads to equation \eqref{trick1}. Thus, we obtain $f(t)=e^{-\frac{b}{2m}t}$ again, but in this case, we only get the exponential energy decay, i.e., $E_1(t)=E_2(t)=E_0e^{-\frac{b}{m}t}$, which is a good approximation of the exact energies \eqref{Eeg}, with $\frac{b}{2m}\ll\omega_0$, only if one is interested in the behavior of the energy on time scales greater than $T_0$. This simplified version of our derivation is suitable for high school students, i.e., from a mathematical point of view, it is at a slightly lower level of difficulty than the method presented in \cite{Kontomaris2024} since there is no need to calculate the time integral of the energy dissipation rate if our trick is used.

\section{Incorporating our approach to teaching}
\label{benefit}

As we mentioned in the introduction, there are two typical approaches to this topic in standard physics textbooks for undergraduate studies:
\begin{itemize}
\item The general exact solution of differential equation \eqref{HOeq} is just given \cite{Resnick10, Young2020university}, without derivation, along with the expression for damped angular frequency. We will refer to this approach as approach A.
\item A solution of the form $x(t)=Ae^{-\frac{t}{2\tau}}\cos(\omega t+\phi)$ is assumed \cite{Waves}, without explaining why the exponential function is taken and why the factor $1/(2\tau)$ in the exponent, then, by direct substitution in \eqref{HOeq}, $\omega$ and $\tau$ are determined for which the assumed solution satisfies \eqref{HOeq}. We will refer to this approach as approach B.   
\end{itemize}
Here, we comment on the benefits of having the instructor first introduce students to our modeling of the solutions and energies of a weakly damped harmonic oscillator and then continue with approach A or approach B.

Through section \ref{model}, students can learn that the amplitude decays exponentially in the weak damping regime and that the relationship between $\frac{b}{2m}$ and $\omega_0$, i.e. between $b$ and $2\sqrt{km}$, characterizes the damping strength, and not the magnitude of $b$ per se. These findings are not given to them as finished products without derivation and without explanation, i.e., students can adopt them with understanding because our approach uses a mathematical formalism with which they are familiar and comprehensible approximations. Furthermore, instructors can present figures with our modeled energies, as Fig.\,\ref{Eusporedba}, without comparing them to exact energies. By considering such figures, students can connect plateaus in modeled energies with the turning points, and the importance of the monotonous decrease of energy can be pointed out to them, i.e., the energy of free-damped oscillations cannot increase. This last point is sometimes overlooked even by experienced authors, e.g., in \cite{Kontomaris2024,Dourmashkin} shown examples of exact energies of free damped harmonic oscillator display visible local maxima, which immediately indicates some plotting error, and can be potentially confusing to students. Before continuing with approach A or B, an additional helpful step would be to show a figure such as Fig.\,\ref{xusporedba}. It would be sufficient to show the behavior for initial conditions with a purely potential initial energy for students to see that, for larger values of $\frac{b}{2m\omega_0}$, both frequency and phase begin to deviate from the frequency and phase of the undamped case. That the phase of the damped solution must be changed compared to the phase of the undamped solution is also evident from the discussion in section \ref{model}. 

If the instructor now continues with approach A, i.e., gives students the general exact solution of \eqref{HOeq} and the expression for damped angular frequency, even without derivation, students will have a clear understanding of the exponential function in the exact solution and intuition about the frequency decrease of damped oscillations in comparison to the undamped case. If the instructor continues with the approach B, he (or she) can try a solution of the form 
\begin{equation}
x(t)=Ae^{-\frac{b}{2m}t}\cos(\omega t+\phi)\,,
\label{ansatz}
\end{equation}
and by direct substitution in \eqref{HOeq} find that damped angular frequency has to be $\omega=\sqrt{\omega_0^2-\left(\frac{b}{2m}\right)^2}$, i.e. students will have clear idea why the instructors takes this form of the solution, and there is no need for introducing unknown and unintuitive factor $1/(2\tau)$ in the exponential function. At his point in both scenarios, the instructor can define the damping coefficient $\gamma=\frac{b}{2m}$, for simplicity, and specify that the damped system oscillates only if $\gamma<\omega_0$, i.e. in the underdamped regime, and, depending on the course program, further comment on the critical damping, i.e. $\gamma=\omega_0$, and overdamping, i.e. $\gamma>\omega_0$, regimes.

We note here that after the students are familiar with the solution \eqref{ansatz}, the instructors can obtain the exact solutions of the critical and overdamped regimes without ever formally solving the differential equation \eqref{HOeq}, i.e. the solutions of those two regimes are already contained in solution \eqref{ansatz}. To see that, one proceeds as follows. Solution \eqref{ansatz} can be written as $$x(t)=e^{-\gamma t}\left(B\cos(\omega t)+C\sin(\omega t)\right)\,,$$ and when we determine the constants $B$ and $C$, from initial conditions $x_0$ and $v_0$, we get
\begin{equation}
x(t)=e^{-\gamma t}\left(x_0\cos(\omega t)+\frac{v_0+\gamma x_0}{\omega}\sin(\omega t)\right)\,.
\label{xud}
\end{equation}
We can easily obtain the solution of critical damping regime by taking the $\gamma\rightarrow\omega_0$ limit of \eqref{xud}, since in that case $\omega\rightarrow0$, $\lim_{\omega\rightarrow0}\cos(\omega t)=1$ and $\lim_{\omega\rightarrow0}\omega^{-1}\sin(\omega t)=t$. To obtain the solution of the overdamped regime, one can write $\omega=i\Omega$, where $\Omega=\sqrt{\gamma^2-\omega_0^2}$ and $i=\sqrt{-1}$, and use the connection of trigonometric and hyperbolic functions, i.e. $\sin(i\Omega t)=i\sinh(\Omega t)$ and $\cos(i\Omega t)=\cosh(\Omega t)$, in \eqref{xud}, as was done in \cite{Waves}.        

\section{A short comment on experimental backup for our approach}
\label{experiment}

Fig.\,\ref{x1x2} is one of the central ingredients of our approach. It gives us insight into the approximations we can use for weak damping. If we wanted to obtain the data for such a figure experimentally, the only tricky part would be to realize two different pairs of initial conditions with the same initial energy, one with purely potential initial energy and the other with purely kinetic initial energy. Here, we comment on only one simple idea to achieve a good approximation of such initial conditions, which even students without prior knowledge of damped oscillators could understand and apply in the laboratory. As a concrete example, we will consider a physical pendulum with linear damping due to eddy currents \cite{Wang, Gonzalez2006}.

In such systems, all damping regimes can be achieved, from weak damping with $\gamma\sim 0.01\omega_0$ \cite{Wang} to overdamped regime with $\gamma\gtrsim\omega_0$ \cite{Gonzalez2006}. Let us imagine that we are considering such a pendulum in the regime of small oscillations and weak damping with $\gamma\sim0.01\omega_0$, which corresponds to the experimental conditions in \cite{Wang}. We displace the pendulum by the angle $\theta_0$, let it swing freely for some time and record the data. This way, we would obtain data similar to the one shown as a red curve in Fig.\,\ref{x1x2}(a), with angle as a displacement. Then, by comparison with the undamped case with the same initial conditions, we could easily see that the experimental data and undamped solution mostly overlap for $0\leq t\leq T_0/2$, with only slightly smaller amplitude of damped displacement at the first turning point, i.e. at $t=T_0/2$. Thus, we could conclude that the energy loss due to damping is negligible in the first quarter of a period, i.e. for $0\leq t\leq T_0/4$. If we then repeat the experiment with initial displacement $-\theta_0$ and plot the angle vs. time data starting from $t=T_0/4$, we would get a curve similar to a blue curve in Fig.\,\ref{x1x2}(b), i.e., a curve which corresponds to initial conditions with purely kinetic initial energy which is negligibly smaller than the initial potential energy in the previous experiment.

\section{Conclusion}
\label{conclusion}

We showed how to obtain an excellent approximation of the solutions and the energy of a weakly damped harmonic oscillator without solving the differential equation of the second order, i.e., equation \eqref{HOeq}, and without any previous insights into the analytical form of the solutions of that equation. The model and trick we introduced are suitable for first-year undergraduates. We commented on a simplified version of our model and derivation, which are more appropriate for high school students. In addition, we gave suggestions on integrating our approach, method, and results into teaching. We commented on obtaining the exact solutions for all damping regimes without using formal methods for solving second-order differential equations.      


\section{Acknowledgments}

This work was supported by the QuantiXLie Center of Excellence, a project co-financed by the Croatian Government and European Union through the European Regional Development Fund, the Competitiveness and Cohesion Operational Programme (Grant No. KK.01.1.1.01.0004).

\bibliographystyle{unsrt}
\bibliography{PublicationList.bib}

\end{document}